\documentclass[12pt,a4paper]{article}
\usepackage{a4,latexsym,amsfonts}
\newcommand{\be}{\begin{equation}}
\newcommand{\ee}{\end{equation}}
\newcommand{\ed}{\end{displaymath}}
\newcommand{\bd}{\begin{displaymath}}
\newcommand{\bea}{\begin{eqnarray}}
\newcommand{\eea}{\end{eqnarray}}
\newcommand{\barr}{\begin{eqnarray}}
\newcommand{\earr}{\end{eqnarray}}
\newcommand{\bas}{\begin{eqnarray*}}
\newcommand{\eas}{\end{eqnarray*}}
\newcommand{\R}{\mbox{$\rho$}}

\newcommand{\br}{\mbox{$\bar \rho$}}
\newcommand{\bz}{\mbox{$\bar z$}}
\newcommand{\om}{\mbox{$\Omega$}}
\newcommand{\lap}{\mbox{$\bigtriangleup$}}
\newcommand{\dx}{\mbox{$\partial$}}
\begin{document}
\title{Asymptotically Flat Initial Data for\\ Gravitational Wave Spacetimes,\\
       Conformal Compactification\\ and Conformal Symmetry}
\author{Sascha Husa\\
{\em Department of Physics and Astronomy, University of Pittsburgh}}
\date{October 11, 1998}
\maketitle

\begin{abstract}
We study the utilization of conformal compactification within the conformal
approach to solving the constraints of general relativity for asymptotically
flat initial data.
After a general discussion of the framework, particular attention is paid to
simplifications that arise when restricting to a class of initial data which
have a certain $U(1)\times U(1)$ conformal symmetry.
\end{abstract}

\section{Introduction}

Applying the conformal method \cite{York_in_Smarr} of solving the constraints
of general relativity to asymptotically flat (AF) initial data, the usual
choice is to pick an AF representative of the conformal equivalence class as
base metric.
Here instead we discuss the alternative of choosing a conformally compactified
seed geometry.
This formalism offers a number of advantages, e.g. it is not necessary to
introduce any artificial cutoff at some finite radius in numerical work,
and {\em all} asymptotically flat regions in multiple-black-hole initial data
can be treated on an equal footing. Furthermore the method is well suited to
exploit a certain conformal symmetry, which yields further simplifications.

The conformal symmetry considered here consists of a $U(1)\times U(1)$ group
of {\em conformal isometries} associated with two commuting, orthogonal
conformal Killing vector fields, and will be called {\em (conformal) toroidal
symmetry}. Their action extends to a $U(1)\times U(1)$ symmetry on the
many-point compactification $S^3$ of the physical initial hypersurface.

Based on the discussion presented here, a numerical analysis, including the
existence and properties of apparent horizons, has been carried out for (i)
Brill waves (plus black holes), (ii) initial data containing a marginally
outer trapped torus,
and (iii) time-asymmetric initial data, where the extrinsic curvature is
obtained as an exact solution for non-conformally flat geometries with
conformal symmetry. These results have been discussed in \cite{thesis} and
\cite{mint}.
The organization of this paper is as follows: first we briefly review the
well known formulation of the constraints on a compactified background in
Sec. \ref{sec:comp_background}, then we discuss toroidal symmetry and the
usual Brill ansatz for axisymmetric initial data in its compactified,
toroidally symmetric incarnation in Sec. \ref{sec:t-sym}.
Section \ref{sec:light} discusses the constraints in the context of toroidal
symmetry, in particular the case of time symmetry, while Sec. \ref{sec:morebhs}
deals with the case of multiple black holes.

 \section{Constraints and Compactification}\label{sec:comp_background}

\subsection{Time Symmetry}

In this section, we will briefly discuss the formulation of the constraints on
a compactified background.
The standard 3+1 split of Einstein's equations yields one vector (momentum)
and one scalar (Hamiltonian) constraint. 
Choosing a maximal slice (defined as a slice of vanishing mean curvature)
as the initial Cauchy surface
$\Sigma$, the Hamiltonian constraint reads
\be\label{HC}
{^{3}R[h_{ab}]}=K_{ab}K^{ab},
\ee
where ${^{3}R[h_{ab}]}$ is the scalar curvature of the spatial 3-metric
$h_{ab}$, and $K_{ab}$ is its extrinsic curvature.
The momentum constraint on a maximal slice is
\be\label{MC}
D_{b}{K^{b}}_{a}=0.
\ee

For simplification the time-symmetric case ($K_{ab}=0$)
is discussed first. Then the momentum constraint (\ref{MC}) is satisfied
identically,
and only the Hamiltonian constraint (\ref{HC})  remains.
In the conformal approach to solving the constraints,
the physical metric is constructed from a freely specifiable seed metric
$\bar h_{ab}$ via conformal rescaling:
$h_{ab} = \psi^{4}\bar h_{ab}$. Assuming $\bar h_{ab}$ to be
asymptotically flat
this ansatz transforms the Hamiltonian constraint into the linear elliptic
equation
\be
\label{HC_conformal_at_Kab=0}
L_{\bar h}\psi\equiv
\left(-\lap_{\bar h}+\frac{1}{8}R_{\bar h}\right)\psi=0,\qquad
\psi>0,\qquad\psi\rightarrow 1{ \mbox{ at } }\infty,
\ee
for the conformal factor $\psi$. The differential operator $L_{\bar h}$ is
called the conformal Laplacian of the metric $\bar h$.

Note that, since the mass can be given an arbitrary value by conformal
rescalings,
we may in particular choose a base metric with zero mass for simplicity.
The mass is then fully encoded in the conformal factor that solves the
Hamiltonian constraint.
For a massless metric $\bar h$, there exist regular one-point
compactifications $(\bar\Sigma,\bar g)$ of $(\Sigma, \bar h)$.
So $\bar g_{ab}=\omega^{2}\bar h_{ab}$ is a smooth metric on
$\Sigma=\bar\Sigma-\Lambda$, where $\Lambda$ is the point ``at infinity'',
it can be extended smoothly to $(\bar\Sigma,\bar g)$ 
and $\omega$ is a regular asymptotic distance function
(RADF) of $(\bar\Sigma,\bar g)$ at $\Lambda$ (compare e.g. 
\cite{geroch_spatial_infinity} and \cite{bobby-RADF}).
A function $\omega$ is called an RADF near a point $\Lambda$ of a manifold
$(\bar\Sigma,\bar g)$ if it is $C^{\infty}$, positive and satisfies
\be\label{def_RADF}
\omega\vert_{\Lambda} = 0, \quad
\bar D_{a}\omega\vert_{\Lambda} = 0, \quad
 (\bar D_{a}\bar D_{b}\omega - 2\bar g_{ab})\vert_{\Lambda} = 0,\quad
 \bar D_{a}\bar D_{b}\bar D_{c}\omega\vert_{\Lambda} = 0.
\ee
In asymptotically Cartesian coordinates on $\Sigma$, the asymptotic behavior
of$\omega$ is given by $\omega=O(r^{-2})$ near infinity.

Going over to the compactified picture now, we consider the equation
\be\label{HC_conformal_c_Kab=0}
\left(-\bigtriangleup_{\bar g}+\frac{1}{8}R_{\bar g}\right)G =
                                                      4\pi\delta(\Lambda),
\ee
where $\delta(\Lambda)$ is the $\delta$-distribution centered at $\Lambda$,
and ${\bar g}_{ab}$ is the regular metric $\bar g_{ab}=\omega^2\bar h_{ab}$
on  $\bar\Sigma$, the one-point compactification of $\Sigma$,.
 
The function $G$ is required to be positive and smooth
on $\bar\Sigma\backslash \Lambda$.
Near $\Lambda$ such a solution will blow up, in particular it is known
\cite{lp} that
\be\label{G asympt}
G=\omega^{-1/2}+\frac{m}{2}+O(\omega^{1/2}),\qquad m=const.,
\ee
where $\omega$ is a RADF near $\Lambda$ and the constant $m$ is independent
of the choice of RADF.

The manifold $(\Sigma,\bar h)$ with $\bar h_{ab}=\omega^{-2} \bar g_{ab}$
is then asymptotically flat near $\Lambda$ with zero ADM mass.
The metric $h_{ab}=\psi^4 \bar h_{ab}=G^4 \bar g_{ab}$ however, due to
Eq. (\ref{G asympt}), is asymptotic to the Schwarzschild metric of mass $m$:
\be\label{psidef}
\psi=\omega^{1/2}G= 1 + \frac{m}{2}\omega^{1/2} + O(\omega),
\ee
and has zero scalar curvature since
$(-\lap_{\bar g}+\frac{1}{8}R_{\bar g})G=0$ on $\Sigma$.
Its mass $m$ is guaranteed to be positive by the positive mass theorem
\cite{pos_mass}.

\subsection{The General Case}\label{sec:generalcase}

We will now incorporate extrinsic curvature into the formalism.
For vanishing trace of the extrinsic curvature (i.e. a maximal slice) the
momentum constraint $D_{a}K^{ab}=0$ is conformally invariant if $K^{ab}$ is
rescaled to $\bar K^{ab}=\psi^{10}K^{ab}$. This holds in
particular if the conformal rescaling corresponds to compactification.
Here we start with a symmetric tensor $P_{ab}$ on the compact base manifold,
which is transverse and trace-free (TT) with respect to the compactified base
metric $\bar g_{ab}$. Then we obtain the TT tensor used in the AF setting by 
rescaling to  $\bar K_{ab}=\omega P_{ab}$,
which enters the Hamiltonian constraint in its usual formulation
\be
\left(-\lap_{\bar h}+\frac{1}{8}R_{\bar h}\right)\psi
=\frac{1}{8}\bar K_{ab}\bar K^{ab}\psi^{-7},
  \qquad\psi\rightarrow 1{ \mbox{ at } }\infty.
\ee
The physical extrinsic curvature is then obtained by the  
rescaling $K_{ab}=G^{-2} P_{ab}$. In order to compute the
asymptotic behavior of $P_{ab}$, we have to take into account both the
rescaling, and the fact that the tensor components will be evaluated in
different coordinate systems, which yields
\be
K_{ij} = \frac{\partial\bar x^{\bar i}}{\partial x^{i}}
       \frac{\partial\bar x^{\bar j}}{\partial x^{j}}
K_{\bar i \bar j}
       = \frac{\partial\bar x^{\bar i}}{\partial x^{i}}
       \frac{\partial\bar x^{\bar j}}{\partial x^{j}}
P_{\bar i \bar j}G^{-2} 
       = P_{\bar i \bar j}\times O(\omega^{3}),
\ee
where $K_{ij}$ are components in an asymptotically Cartesian coordinate system
$x^{i}$, and $P_{\bar i \bar j}$ are components in a
coordinate system $\bar x^{\bar i}$ which is regular on $S^3$ near $\Lambda$.
For asymptotically Schwarzschildian data ($K_{ab}=O(\omega^{3/2})$) we
therefore choose $P_{ab}=O(\omega^{-3/2})$. In the more general case
allowing for linear momentum we have $P_{ab}=O(\omega^{-2})$.

In order to generalize the Hamiltonian constraint to the non-time-symmetric
case, we consider the equation
\be
\left(-\bigtriangleup_{\bar{g}}+\frac{1}{8}R_{\bar g}\right)H =
                    4\pi\delta(\Lambda) + H^{-7}\frac{1}{8}P_{ab}P^{ab}.
\ee
This can be treated as follows: denote by $G$ a solution to the
time-symmetric Hamiltonian constraint and define
\bd
H= G + \phi.
\ed
This yields the following equation for the function $\phi$
\be\label{HC_for_h}
L_{\bar g}\phi=\frac{1}{8}(1+G^{-1}\phi)^{-7}G^{-7}P_{ab}P^{ab}.
\ee
Note that this equation is regular for Schwarzschildian initial data, since
then
$G^{-7}P_{ab}P^{ab}=O(\omega^{1/2})$.
It follows from the Appendix of \cite{thefamousappendix} that there exists a
unique positive solution $H$ of Eq. (\ref{HC_for_h}), smooth on $\Sigma -
\Lambda$, bounded and with bounded first derivatives on $M$. 

Existence and uniqueness of solutions to the
Hamiltonian constraint for given $(P_{ab},\bar g_{ab})$ can be inferred from
the time-symmetric case.
It is known that a unique positive solution for
(\ref{HC_conformal_c_Kab=0}) exists, iff the Yamabe number \cite{yamabe}
of $\bar g_{ab}$
is positive, or, equivalently if  the lowest eigenvalue of
$L_{\bar g}$ is positive. This result was effectively proven by Cantor and
Brill \cite{cantor}.
Existence can thus be read off from  the solution of the {\em linear} equation
\begin{equation}\label{lambda_eigenvalue}
L_{\bar g}\theta=\lambda\,\theta,
\end{equation}
and is in particular independent of the choices of $P_{ab}$ and $\Lambda$.

\section{Toroidal Conformal Symmetry and the Brill Ansatz}\label{sec:t-sym}
 
In this section we will study metrics on $S^{3}$ in the context of toroidal
symmetry.
To start we write the standard metric on $S^3$, the one-point compactification
of $R^3$, in toroidal coordinates (see e.g. \cite{tid}) as
\be
ds^2=\frac{d\br^{2}}{1-\br^{2}}+\left(1-\br^{2}\right)d\varphi^{2}
+\br^{2}d\bz^{2},
\ee
where $0\leq \bar\rho\leq 1$, $0\leq\bar z, \varphi\leq 2\pi$.
The vector fields $(\partial/\partial\varphi)^{a}$, $(\partial /
\partial\bar z)^{a}$ form a
pair of commuting, orthogonal and hypersurface-orthogonal Killing vector
fields of $g$ spanning the surfaces of constant $\br$. For $\br
\neq 0,1$ these are flat tori, as can be seen immediately by
inspection of the metric. 
Decompactifying to the flat metric by stereographic projection
a torus $\br=const.$ is mapped to a standard (non-flat) torus,
for $\varphi=const.$.
The sets $\br=0$ (respectively $\br=1$) are linked great circles
on $S^{3}$, corresponding to the $z$-axis
(respectively the circle $z=0,r=1$)
after stereographic projection. For more details see e.g. \cite{tid},
\cite{mint} or \cite{thesis}.

In order to achieve a convenient form of the Hamiltonian constraint
for axially symmetric initial data, we use the well known Brill ansatz
\cite{brill}, which is usually formulated in the asymptotically flat picture,
but straightforwardly translates to                           
\begin{eqnarray}
h&=& \psi^{4}\left(e^{2a q\left(\rho,z\right)}\left(d\R^{2}
+dz^{2}\right)+\R^{2}d\varphi^{2}\right)\nonumber\\[0.5cm]
&=& G^4 \left(e^{2a q\left(\bar{\rho},
\bar z\right)}\left(\frac{d\br^{2}}{1-\br^{2}}
+\left(1- \br^{2}\right)d\bz^{2}\right)+\br^{2}d\varphi^{2}\right)
=\psi^{4}\Omega^{-2}\bar g.
\end{eqnarray}
Here we have chosen the conformal factor for compactification as
$\omega=\Omega=2/(1+r^2)$, which corresponds to stereographic projection,
and is compatible with the symmetries we want to exploit. Note that
$\Omega$ is actually not an RADF, but $2 \Omega$ is.
For the base metric $\bar g$ we can now read off the ansatz 
\be\label{brillkomp}
{\bar g} = e^{2a q\left(\bar{\rho},\bar z\right)}
\left(\frac{d\br^{2}}{1- \br^{2}}
+\left(1- \br^{2}\right)d\bz^{2}\right) + \br^{2}d\varphi^{2}.
\end{equation}

The crucial step is to choose $q=q(\br)$. This implies that $(\partial/
\partial\bz)^{a}$ is a Killing vector field (KVF) of $\bar g_{ab}$
in addition to axial symmetry.
At the axis $\bar\rho=0$ the two conditions $q(0)=0$,
$q'(0)=0$ are required to ensure both regularity on the axis and the
correct asymptotic behavior (see \cite{brill} and \cite{thesis}).
A deformation of the standard metric on $S^3$ with $q=q(\br)$ now respects
both KVF's $(\partial/\partial\varphi)^{a}$ and
           $(\partial/\partial\bar z )^{a}$.
The scalar curvature simplifies to
\be
\frac{R_{\bar g}}{8}(\bar\rho)=e^{-2aq}\left(\frac{3}{4} + V(\bar\rho)\right)
\ee
where $V(\bar\rho)= -\frac{a}{4}\left[\left(1-\br^{2}\right)q''
- 2\br q'\right]$. The action of the conformal
Laplacian on axially symmetric functions is thus given by
\be
L_{\bar g} f(\br,\bz)=e^{-2aq\left(\bar\rho\right)}\left(-\lap_{0}+\frac{3}{4}
 + V\right)f.
\ee
This simplification will be the key to turn PDEs into ODEs in the next
section.

 \section{The Constraints in the Light of Conformal
          Symmetry}\label{sec:light}

  \subsection{Existence of Solutions to The Hamiltonian Constraint }

Following Sec. \ref{sec:generalcase}, to determine the existence of a
solution to the Hamiltonian constraint is equivalent to
positivity of the lowest eigenvalue $\lambda_{1}$ of $L_{\bar g}$.
Assuming toroidal symmetry, the eigenvalue problem
\begin{displaymath}
L_{\bar g}\theta = \lambda\theta,
\end{displaymath}
is separable (since $q=q(\br)$, hence the potential depends on $\br$
only) and the solutions can be written as
\begin{displaymath}
\theta_{\lambda m n}(\br,\varphi,\bar z)=\vartheta_{\lambda m n}
(\br)\sin(n\varphi + \alpha)\sin(m\bar z + \beta),
\end{displaymath}
where $\alpha$ and $\beta$ are arbitrary angles. We are thus left to solve an 
equation for $\vartheta$,
\begin{displaymath}
\left(L_{\br}
      + \frac{m^2}{1- \br^{2}}
      + \frac{n^2}{\br^{2}}\right)\vartheta_{\lambda m n} =
                \lambda\vartheta_{\lambda m n},
\end{displaymath}
where $L_{\br}$ is the {\em radial} part of the conformal Laplacian
$L_{\bar g}$,
\begin{equation}\label{L_reduced_light}
L_{\br}=-(1- \br^{2})\frac{\partial^2}
   {\partial \br^{2}}
      - \frac{1-3 \br^{2}}{\br}\frac{\partial}{\partial \br}
      + \frac{3}{4}  + V(\br).
\end{equation}
Since the first eigenfunction is non-degenerate \cite{kazdan} and has
therefore to depend trivially on the angular coordinates $\varphi$ and
$\bar z$, we may set $m=n=0$ when looking for the lowest eigenvalue
$\lambda_{1}:$
\begin{equation}\label{existence-ODE-light}
L_{\br}\vartheta_{\lambda 00} = 
             \lambda\vartheta_{\lambda 00}.
\end{equation}

The existence question has thus been reduced to an ODE of Sturm-Liouville
type which can be solved by standard methods.

\subsection{The Time-Symmetric Hamiltonian
Constraint}\label{sec:light:time-symm}

We will now use the setup developed in Sec. (\ref{sec:comp_background}) 
to convert the time-symmetric Hamiltonian constraint into a set of uncoupled
ODEs.
Using $q(\Lambda)=0$, so that $\exp(2aq)\delta(\lambda)\equiv\delta(\Lambda)$,
the constraint equation reads
\be\label{comp-constraint.explicit}
\left( -\lap_{0}+\frac{3}{4}+V\right)G(\br,\bz) =
 4\pi\,\sqrt{2}\delta\left(\Lambda\right),
\ee
where the factor of $\sqrt{2}$ was introduced to compensate for
$\om$ not being an ADF, so that the solution of the unperturbed ($a=0$)
case becomes $G_{0}=\om^{-1/2}$. Nevertheless we will sloppily
call $G$ a Green function.

Before we actually transform (\ref{comp-constraint.explicit}) to ODEs, we
will formulate it as an equation for a regular function instead of being
an equation for the singular $G$, whose asymptotic behavior near $\Lambda$
was given in (\ref{G asympt}).
To remove the distributional character of this equation we write
the solution as a perturbation of the Green function of the conformally flat
metric:
\be
G = G_{0} + \phi
\ee
and get an equation for the continuous function $\phi$:
\be
\left( -\lap_{0} + \frac{3}{4}+V \right) \phi = -V G_{0}.
\ee
Depending on the choice of $q$ the inhomogeneous term $V G_{0}$ still may show
a singular behavior. If we have $q=O(\br^m)$ then $V=O(\br^{m-2})$ and
\bd
V G_{0} = O(\frac{\br^{m-2}}{\sqrt{\bz^2 + \br^2}})
\ed
which is singular for $m\leq 3$. Later we will cosine transform with respect
to $\bz$ and thus have to consider finiteness of the integral
\bd
\int V G_{0} \,\, d\bz
\ed
which will be finite for $m > 2$ ($V_0:=V(0)=0$).
The boundary condition $q=O(\br^2)$ translates to $m\geq 2$,
so $m=2$ is both the minimal value allowed by the boundary condition and
the critical value for the finiteness of the Fourier transform of
the inhomogeneous term. Note that for the critical value the divergence is
logarithmic. To further regularize the equation for $m=2$ we make the ansatz
\be\label{ansatz}
G=\Omega^{-1/2} + c_1 \Omega^{1/2} + c_2 \Omega^{3/2} + \phi.
\ee
Now the constants $c_1$, $c_2$ are chosen so that the contributions of the
$\Omega^{-1/2}$ and $\Omega^{1/2}$ terms cancel at $\Lambda$.
The necessary choices are
\bd
c_1=V_0 ,  \qquad c_2 = \frac{1}{6}V_0(2+V_0).
\ed
The equation for $\phi$ then becomes
\be\label{theq}
\left(\left(1-\br^2\right)\frac{\partial^2}{\partial \br^2}
+\frac{1-3\br^2}{\br}\frac{\partial^2}{\partial \br^2}
+\frac{1}{1-\br^2}\frac{\partial^2}{\partial \bz^2}
-\frac{3}{4}-V \right) \phi = F(\br, \bz),
\ee 
where the function $F$ is given by
\be
F = \left(V - V_0 \right) \om^{-1/2}
+V_0 \left(V - V_0 \right) \om^{1/2}
+\frac{V_0}{6} \left(2 + V_0 \right)
\left(6 + V\right)\om^{3/2},
\ee
which reduces to
\be
F = V \,\Omega^{-1/2}
\ee
for $q = O(\br^{2+m})$.

The simple form of (\ref{theq}), namely that the conformal Laplacian does not 
explicitly depend on $\bz$, suggests the method of expanding
$\phi (\br ,\bz )$ into an even Fourier (cosine) series ($\phi$ is invariant
under reflection at the equator)
\be
\phi (\br ,\bz ) = \sum_{n=0}^{\infty} \phi_{n}(\br) \cos{n\bz}.
\ee
Applying an analogous transformation to $F$ gives
a sequence of uncoupled linear ordinary differential equations
for the Fourier coefficient functions $\phi_{n}$:
\be\label{basic-ode}
\left(\left(1-\br^2\right)\frac{\partial^2}{\partial \br^2}
+\frac{1-3\br^2}{\br}\frac{\partial}{\partial \br}
-\frac{n^2}{1-\br^2} -\frac{3}{4}-V\right) \phi_{n}(\br) = F_{n}(\br).
\ee
Equation (\ref{basic-ode}) has singular points at $\br=0,1$, which can be
associated with
the two axes of symmetry and the corresponding coordinate singularities.
Boundary conditions are necessary to guarantee that the solutions are regular.

The axis at $\br=0$ is also a physical axis of symmetry, and the boundary
conditions there directly correspond to the ones in the asymptotically flat
context.
In addition to requiring
regularity of $q$ and the  conformal factor $\psi$ ($q,$ $\psi$
$\in C^{2}({\mathbb R}^2)$, this implies $q_{\rho}$ and $\psi_{\rho}$ both
have
to vanish on the axis. From (\ref{ansatz}) and the definition
of $\psi$ as $\psi=\Omega^{1/2}G$ (where $\Omega$ is the compactifying
conformal factor) we get
\bd
\psi(\br(x,y,z),\bz(x,y,z)) = 1 + V_{0}\,\Omega + \frac{V_{0}}{6}
\left( 2 + V_0 \right)\Omega^{2} + \om^{1/2}\phi
\ed
so that instead of working with $\psi$ we impose equivalent conditions
on $\phi(\br(x,y,z)$, $\bz(x,y,z))$, which is simply
the sum of the Fourier terms $T_n = \phi_n(\br) \cos{n\bz}$.
For $\br<1$ the coordinate transformation
from cylindrical coordinates to toroidal coordinates is regular, so we simply
require $\phi_n(\br)$ to be a regular function in this interval with
$\phi'=0$ at $\br=0$.

By contrast the line $\br=1$ is only an axis of symmetry for the compact
manifold and the coordinate singularity there does not bear any physical
meaning, and indeed the situation at $\br=1$ is more delicate.
The coordinate transformation from
$(\R, z)$ to $(\br, \bz)$ is singular there.
The boundary conditions on the functions $\phi_n(\br)$
at $\br=1$ have to ensure that the nth Fourier term
$T_n =\phi_n(\br)\cos{n\bz}$ is a regular function of the coordinates
$x$, $y$, $z$.
One can show that (see \cite{thesis})
a necessary and sufficient condition for regularity of $T_n$ is that
\bd
\varphi_{n}(\br) = \frac{\phi_{n}(\br)}{\sqrt{1-\br^2}^n}
\ed
is a regular function.

We also introduce, analogously to $\varphi_{n}$, the functions
\bd
f_{n}=\sqrt{1-\br^2}^{-n}F_{n}.
\ed
We get
\be\label{eq for varphi}
\left(\left(1-\br^2\right)\frac{\dx^2}{\dx \br^2}
+\frac{1-(3+2n)\br^2}{\br}\frac{\dx }{\dx \br}
-n^2 - 2n - \frac{3}{4} - V \right) \varphi_{n} = f_{n}.
\ee

The boundary value problem of finding a positive Green
function has thus been translated into the problem to find
$C^{2}$ solutions to (\ref{eq for varphi}) for every $n$. An
approximate solution $G_N$ is constructed by solving for $n\leq
N$, resulting in
\be
G_N = \Omega^{-1/2} + V_{0}\Omega^{1/2} + \frac{V_{0}}{6}
\left( 2 + V_0 \right)\Omega^{3/2} + \sum_{n=0}^{N}\sqrt{1-\br^2}^{n}
\varphi_{n}\cos{n\bz}.
\ee
An efficient method for solving one-dimensional boundary
value problems is the shooting and matching technique, which has been
used in the numerical study discussed in \cite{thesis}.

So far we have only explicitly dealt with the axially symmetric case, by
assuming that the $\delta$-distribution source was placed on the axis
$\br=0$, in particular at $\bz=0$. This renders $G$ and $\phi$ axially
symmetric and in the Fourier series we only have to consider the 
$\cos n\bz$ terms.   
The generalization of the axially symmetric to the general 3D case is
straightforward. The only places where axial symmetry has been used so far
is in the placing of the source, and the corresponding functional form of
the Green function of the conformally flat conformal
Laplacian $G_{0}(x,\Lambda)$, where $\Lambda$ is placed on the axis
$\rho=0$, which is
$$
G_{0}= (1-\sqrt{1-\br^2}\cos{\bz}).
$$
For a general position of the source at $(\br'$, $\bz'$, $\varphi')$ we
instead have that
$$
G_{0}= (1-\sqrt{1-\br^2}\sqrt{1-\br'^2}\cos{(\bz-\bz')}-\br\br'
   \cos{(\varphi-\varphi')}).
$$
This can again be Fourier expanded in a double Fourier series in $\bz$ and
$\varphi$.

Note that the structure of KVF's of the physical metric depends on
the choice of ``point at infinity''. The condition has been formulated as a
theorem by Beig \cite{CKV}:
A KVF $\eta^{a}$ of a metric $g_{ab}$ is {\em also} a KVF of the
decompactified metric $G^{4}g_{ab}$, where $G$ satisfies the time symmetric
Hamiltonian constraint, $L_{g}G=4\pi\delta{\Lambda}$, with $\Lambda$ the
point at infinity if and only if
$\Lambda$ is a fixed point, i.e. $\eta^a \vert_{\Lambda}=0$.

Since the axes will turn out to be disjoint in the present case,
this means that only one, or even none, of the KVF's may survive
decompactification.

\section{Initial Data Containing Black Holes}\label{sec:morebhs}

The formalism of solving the constraints on a conformally compactified
background discussed in Sec. \ref{sec:comp_background} and
the simplifications due to assuming conformal symmetry described in Sec.
\ref{sec:light} can be immediately generalized to the case when there
are several asymptotic ends. 
We simply choose a finite number of points $\Lambda_{i}$ $(i=1,\dots,N)$,
solve numerically for the corresponding Green function $G_{i}$
$$
L_{\bar g}G_{i}=4 \pi c_{i} \delta(\Lambda_{i}),
$$
and define
\begin{equation}\label{superposition}
h_{ab}=\left(\sum_{i=1}^{N}c_{i}G_{i}\right)^{4}\bar g_{ab},
\end{equation}
where the ``source strengths'' $c_{i}$ are arbitrary positive numbers. When
the amplitude parameter $a$ vanishes, these are simply the many-black-hole
solutions discussed  by Brill and Lindquist \cite{brlind}.   
The existence of the $G_{i}$ is guaranteed by a positive Yamabe number of
$\bar g$. Thus, starting from regular initial data (with positive Yamabe
number), we can always add as many black holes as we like, with arbitrary
weights $c_{i}$ and positions $\Lambda_{i}$.
While the Green function yields the solution to the time-symmetric constraint,
the generalization to incorporating extrinsic curvature is again
straightforward along the lines of Sec. \ref{sec:generalcase}, in particular
one ends up with Eq. (\ref{HC_for_h}). A similar approach to treat
black-hole initial data, which does {\em not} use conformal compactification,
has recently been discussed by Brandt and Br\"ugmann \cite{punctures}
in the conformally flat case.

If all the $\Lambda_{i}$ lie on the two axes,
each $G_{i}$ in $G=\sum_{i=1}^{N}c_{i}G_{i}$ corresponds to an
axially symmetric (Brill-type) solution.
If at least one of the sources is placed off-axis, the resulting physical
geometry will not have any Killing vector fields at all.

Starting with a globally regular Brill-wave solution and adding sources, it is
interesting to observe that Brill's simple positive-mass proof \cite{brill}
immediately gives positivity of mass also in this non-axially symmetric,
topologically nontrivial situation:
For a single source placed at $\Lambda_{1}$ we have that 
$$
G_{1}=G_{0} + \frac{m_{1}}{2} + O(G_{0}^{-1}),
$$
where positivity of $m_{1}$ is guaranteed by Brill's argument \cite{brill}.
Adding sources we get a different mass $\bar m_{1}$ at $\Lambda_{1}$,
$$
\bar m_{1}=2(G - G_{0})\vert_{\Lambda_{1}} = m_{1} +
2\sum_{n=2}^{N} G_{i}\vert{\Lambda_{1}},
$$ 
where the second term is positive by positivity of each individual $G_{i}$.

Although $\partial/\partial\bz$ is {\em not} a KVF of the physical metric
$h_{ab}=G^4 \bar g_{ab}$, i.e. arbitrary rotations in $\bz$ are not
symmetry operations, $h_{ab}$ will possess a discrete symmetry of
invariance under coordinate transformations
$\bz\rightarrow\bz+\delta\bz$ if the $\Lambda_{i}$ are placed equally spaced at
\be
\bz_{i}=\frac{2\pi i}{N+1}, \quad \delta\bz=\frac{2\pi}{N+1}
\ee
and are all equally weighted, that is $w_{i}=1$ with the choice $w_{0}=1$
or more generally for any arrangement of sources that is periodic in $\bz$.

This can be considered a
generalization of inversion symmetry for two asymptotically flat sheets which
corresponds to placing sources of equal strength at antipodal positions.
Like any symmetry of the initial data \cite{Chrusciel-Symmetry} this
symmetry is respected by time evolution.
A nice consequence is                        
that the symmetry carries over to minimal surfaces, that is if one finds one
minimal surface, one can get another $N-1$ minimal surfaces by shifting the
coordinate $\bz$. In particular every asymptotically flat region will show
an apparent horizon. This means that for every choice of the function $q$
the string of black holes will be surrounded by a common apparent horizon.

A particular useful formulation of toroidal symmetry is that, in the case of a
general distribution of $\Lambda_{i}$'s, not all of the $G_{i}$'s have to be
computed separately:
for any $\Lambda_{1}$, $\Lambda_{2}$ lying on the same orbit of the isometry
group, $G_{2}$ is given in terms of $G_{1}$ by a product of suitable
rotations in $\varphi$ and $\bz$.
Take e.g. $N=2$ with $\Lambda_{1}$ at $(\rho=0,\bz=0)$ and $\Lambda_{2}$ at
$(\rho=0,\bz=\alpha)$.

In the special case $\alpha=\pi$, the physical metric
$h_{ab}$ has the mirror symmetry $\bz\mapsto2\pi-\bz$. When $c_{1}=c_{2}$,
it has the inversion symmetry
$\{\bz\}\cup\{\bz+\pi\}\mapsto\{\alpha-\bz\}\cup\{\alpha-\bz+\pi\}$,
 leaving
the ``throat'' $\{\bz=\alpha/2\}\cup\{\alpha/2+\pi\}$ invariant. Thus the
``throat'' is totally geodesic with respect to $h_{ab}$, in particular a
minimal embedded 2-sphere (a ``horizon''). When $h_{ab}$ is conformally flat,
i.e. $a=0$, the
above discrete symmetries are present irrespective of $c_{1}$, $c_{2}$ and
$\alpha$. Because, then, by combining a homothety with a proper conformal
motion coming from the Killing vector $\frac{\partial}{\partial z}$ on $(R^{3},
\bar\delta_{ab})$,
we can always arrange for $c_{1}'=c_{2}'$, $\alpha'=\pi$. The corresponding
physical data is of course nothing but a time-symmetric slice of the Kruskal
spacetime with mass
$m(c_{1},c_{2},\alpha)=2\sqrt{2}c_{1}c_{2}(1-\cos{\alpha})^{-1/2}$.
When $a\not=0$ our solutions for $c_{1}\not=c_{2}$ are not
inversion symmetric. Placing sources of equal strength at $\rho=0$,
$\alpha_{i}=\frac{2\pi i}{N}$, $i=0,1,\dots,N-1$, we find in an analogous
manner that on $R^{3}\backslash\bigcup_{i=1}^{n}\Lambda_{i}$ each asymptotic
end is surrounded by a throat. Thus, viewing $\Lambda_{1}$ as ``infinity'' and the
other $\Lambda_{i}$'s as ``black holes'' or ``particles'' we can in particular
say that these objects, viewed from infinity, are so close together that,
in addition to a horizon for each of them individually, there is one
surrounding them all.

\subsection*{Acknowledgements}
Much of the work presented here is based on ideas of, or was developed jointly
with, my thesis advisor R. Beig, to whom I am also indebted for his support
and encouragement. Thanks to C. Stanciulescu for critically reading and
identifying some errors in the manuscript.
Part of this work was financially supported by Fonds zur F\"orderung der
Wissenschaftlichen Forschung, Project No. P09376-PHY.

%
%
%


\begin{thebibliography}{42}
\bibitem{pos_mass}
For an overview on the positive mass theorem of Schoen and Yau \cite{sy}
and its relation to the Yamabe problem see Ref. \cite{lp}.
\end{thebibliography}
\end{document}